\documentclass[12pt]{iopart}

\usepackage{iopams}  
\begin{document}

\title{Time-Varying Gravitomagnetism}

\author{Bahram Mashhoon}

\address{Department of Physics and Astronomy, University of Missouri-Columbia, Columbia, MO 65211, USA}
\ead{mashhoonb@missouri.edu}
\begin{abstract} Time-varying gravitomagnetic fields are considered within the linear post-Newtonian approach to general relativity. A simple model is developed in which the gravitomagnetic field of a localized mass-energy current varies linearly with time. The implications of this temporal variation of the source for the precession of test gyroscopes and the motion of null rays are briefly discussed.
\end{abstract}

\maketitle

\section{Introduction}\label{s:1}

Astronomical sources, such as neutron stars, generally rotate, but their rotation rates are seldom uniform. For instance, the Earth's rate of rotation decreases very slowly due mainly to tidal friction; in fact, in this process the Earth loses angular momentum, which is then transferred to the orbital motion of the Moon around the Earth. The temporal variation of the proper angular momentum of a massive body in turn generates a time-varying gravitomagnetic field in accordance with the general theory of relativity. The purpose of this paper is to study gravitomagnetic fields generated by the intrinsic temporal variability of sources; this issue was ignored in a previous work on general gravitomagnetic fields of spinning masses~\cite{1}. Only a beginning is made here, since the detailed study of the gravitomagnetic field resulting from a realistic model of a variable source is beyond the scope of this first analysis.

The linear approximation scheme of general relativity is adopted throughout this paper. Thus the spacetime metric tensor $g_{\mu \nu}$ is expressed as $g_{\mu\nu}=\eta_{\mu\nu}+h_{\mu\nu}$, where $\eta_{\mu\nu}$ is the Minkowski metric tensor with signature $+2$ and $h_{\mu\nu}$ is a first-order (``weak") perturbation. The choice of the background global inertial coordinates $x^\mu=(ct,\mathbf{r})$ is not unique; under a slight transformation of coordinates $x^\mu\mapsto x^\mu -\epsilon ^\mu$, we have $h_{\mu\nu }\mapsto h_{\mu\nu}+\epsilon _{\mu,\nu}+\epsilon _{\nu ,\mu}$, so that the gravitational potentials are gauge-dependent quantities. However, the Riemann, Ricci and Einstein curvature tensors are gauge invariant. In particular the Einstein tensor may be written as
\begin{equation}\label{eq:1} G_{\mu\nu}=-\frac{1}{2} \Box \bar{h} _{\mu\nu},\end{equation}
where the trace-reversed potentials $\bar{h}_{\mu\nu}$ are given by $\bar{h}_{\mu\nu}=h_{\mu\nu}-\frac{1}{2}\eta_{\mu\nu}h$. Here $h=\tr (h_{\mu\nu})$, $\Box =\eta^{\alpha \beta}\partial_\alpha\partial_\beta$ and the transverse gauge condition
\begin{equation}\label{eq:2} \bar{h}^{\mu\nu}_{\;\;\;\; ,\nu}=0\end{equation}
has been imposed. The gravitational field equations, $G_{\mu\nu}=(8\pi G/c^4)T_{\mu\nu}$, can then be expressed as
\begin{equation}\label{eq:3} \Box \bar{h}_{\mu\nu}=-\frac{16\pi G}{c^4}T_{\mu\nu}.\end{equation}
The solution of this linear inhomogeneous equation is a superposition of the retarded solution
\begin{equation}\label{eq:4} \bar{h}_{\mu\nu}=\frac{4G}{c^4}\int\frac{T_{\mu\nu} (ct-|\mathbf{r}-\mathbf{r}'|,\mathbf{r}')}{|\mathbf{r}-\mathbf{r}'|} d^3r'\end{equation}
plus the general solution of the homogeneous wave equation. The latter is ignored in this analysis and we simply focus on Eq.~\eref{eq:4} in what follows. Given a consistent model for the source $T_{\mu\nu}$, $\partial_\nu T^{\mu\nu}=0$, $\bar{h}_{\mu\nu}$ is thus determined and then \begin{equation}\label{eq:5} h_{\mu\nu}=\bar{h}_{\mu\nu}-\frac{1}{2}\eta_{\mu\nu}\bar{h},\end{equation}
where $\bar{h}=\tr (\bar{h}_{\mu\nu}$).

It is now necessary to introduce the slow-motion assumption that is needed in our linear post-Newtonian approach to gravitoelectromagnetism (GEM). We require that all motions within the source take place with speeds that are very small compared to $c$. Furthermore, all contributions to the metric tensor of  $O(c^{-4})$ are neglected. It follows from Eq.~\eref{eq:4} that for the sources under consideration here, $\bar{h}_{00}=4\Phi /c^2$, $\bar{h}_{0i}=-2A_i/c^2$ and $\bar{h}_{ij}=O(c^{-4})$. In fact, $\Phi (t,\mathbf{r})$ is the gravitoelectric potential and $\mathbf{A}(t,\mathbf{r})$ is the gravitomagnetic vector potential such that
\begin{equation}\label{eq:6} \frac{2}{c} \frac{\partial \Phi}{\partial t} +\boldsymbol{\nabla} \cdot \mathbf{A}=0,\end{equation}
and the metric is given by
\begin{equation}\label{eq:7} ds^2=-c^2 \left( 1-\frac{2\Phi}{c^2} \right) dt^2 -\frac{4}{c} (\mathbf{A} \cdot d\mathbf{r})dt +\left( 1+\frac{2\Phi}{c^2}\right) \delta_{ij} dx^idx^j.\end{equation}
Our definition of GEM potentials is based on a special convention that preserves the electromagnetic analogy as much as possible~\cite{2}. Moreover, it is useful to define $A^\mu =(2\Phi ,\mathbf{A})$, so that $\bar{h} ^{0\mu}=2A^\mu /c^2$, and the GEM Faraday tensor is then given by $F_{\mu\nu} =A_{\nu,\mu}-A_{\mu,\nu}$. It follows from these defintions and Eq.~\eref{eq:3} that the Maxwell equations for the GEM fields are satisfied~\cite{2}; that is, $F_{\mu\nu}\mapsto (2\mathbf{E},\mathbf{B})$, $F_{[\mu\nu,\rho ]}=0$ and
\begin{equation}\label{eq:8} F^{\mu\nu}_{\;\;\;\; ,\nu}=\frac{8\pi G}{c} j^\mu ,\end{equation}
where $j^\mu$ is the mass-energy current such that
\begin{equation}\label{eq:9} cj^\mu=(T^{00}, T^{0i}).\end{equation}
Thus in the GEM approach adopted in this paper, $T^{ij}$ does not explicitly appear; however, its existence is implicitly assumed such that the four dynamical equations
\begin{equation}\label{eq:10} T^{\mu\nu}_{\;\;\;\; ,\nu}=0\end{equation}
are always satisfied. The linear perturbation approach to GEM has been reviewed in~\cite{3}. The compact notation of Eqs.~\eref{eq:8} and \eref{eq:9} has been employed for simplicity, so that the tensorial character of these equations is purely formal; that is, the background global inertial frame is essentially fixed in this linear GEM treatment.

To study time-dependent gravitomagnetic fields caused by the intrinsic variability of the source, it would be necessary to have a consistent model of a source with time-varying proper angular momentum. This is rather complicated in general; however, a drastic simplification occurs within the linear GEM framework via a simple toy model that is the subject of the next section.

\section{Toy model}\label{s:2}

It is an immediate consequence of Eqs.~\eref{eq:4} and \eref{eq:5} that if $T_{\mu\nu}$ is independent of time, then $h_{\mu\nu}$ is only a function of spatial variables and hence the spacetime of a time-independent source is in general stationary. Let $T^\ast_{\mu\nu}$ represent the energy-momentum tensor of a time-independent source and $\bar{h}^\ast_{\mu\nu}$ be the corresponding stationary solution of Eq.~\eref{eq:4}; then, the time-independent GEM potentials are $\Phi^\ast (\mathbf{r})$ and $A^\ast (\mathbf{r})$, while $\bar{h}^\ast_{ij} =O(c^{-4})$. A detailed treatment of such configurations is contained in~\cite{4}. For time-varying GEM potentials, a simple ansatz would involve a separation of temporal and spatial variables. Under what conditions can
\begin{equation}\label{eq:11} \Phi (t,\mathbf{r})=\varphi (t)\Phi^\ast (\mathbf{r}),\quad\mathbf{A}(t,\mathbf{r})=f(t)A^\ast (\mathbf{r})\end{equation}
represent acceptable GEM potentials? Equation~\eref{eq:6} implies that $d\varphi /dt=0$, while we find from direct substitution of Eq.~\eref{eq:11} in Eq.~\eref{eq:3} that $d^2f/dt^2=0$ and
\begin{equation}\label{eq:12} T_{00}=\varphi T^\ast_{00},\quad T_{0i} =f(t)T^\ast _{0i}.\end{equation}
It follows that our ansatz~\eref{eq:11} is consistent with GEM equations provided that $\varphi$ is constant and $f(t)$ is a linear function of time such that $2 |\mathbf{A}| << c^2$; moreover, the corresponding $T_{ij},\partial _jT^j_i =(df/cdt)T^\ast _{0i}$, is expected to lead to $\bar{h}_{ij}=O(c^{-4})$ via Eq.~\eref{eq:4}.

The mass-energy current~\eref{eq:9} for the case of Eq.~\eref{eq:12} is such that $\mathbf{j}$, $cj^i=T^{0i}$, is localized and divergenceless; therefore, its volume integral vanishes. This implies that Eq.~\eref{eq:11} is consistent with the retarded solution \eref{eq:4}, since $f(t)$ is just a linear function of time. In addition, stresses $T_{ij}$ exist such that the source has proper dynamics in accordance with Eq.~\eref{eq:10}; however, the GEM metric \eref{eq:7} depends only on $\Phi$ and $\mathbf{A}$ and hence the nature of the stresses is of no consequence here. Specifically, these stresses may turn out to be physically unrealistic; however, this is mitigated by the circumstance that the source leads to a simple analytic solution that can be consistently employed within the linear GEM framework. This is illustrated by the special example that is presented in the following section.

With $\varphi$ constant and $f(t)$ linear in time, the GEM fields are
\begin{equation}\label{eq:13} \mathbf{E} (\mathbf{r})=-\varphi \boldsymbol{\nabla} \Phi^\ast -\frac{1}{2c}\frac{df}{dt}\mathbf{A}^\ast,\quad \mathbf{B}(t,\mathbf{r})=f(t)\mathbf{\nabla}\times \mathbf{A}^\ast,\end{equation}
so that $\mathbf{E}$ is independent of time and only $\mathbf{B}$ is linearly dependent upon time. Thus the gravitational analogue of the displacement current vanishes in this case, but our simple model is ideally suited to help elucidate the gravitational analogue of the Faraday law of induction~\cite{5}. Moreover, the linear dependence of the source upon time is consistent with the absence of gravitational radiation in this case.

Many significant relativistic effects of astrophysical interest fall within the linear GEM scheme; therefore, our simple model can provide useful results. In practice, the slow temporal variation of the angular momentum of the source may not be linear; however, our linear model could furnish a first approximation. Some applications of our model are discussed in the rest of this paper using a rather special nonstationary spacetime.

\section{Special nonstationary spacetime}\label{s:3}

Far from any localized stationary source of mass $M$ and angular momentum $J^\ast$, the spacetime metric can be expressed in the form \eref{eq:7}, where the dominant contributions to the GEM potentials are given by
\begin{equation}\label{eq:14} \Phi^\ast =\frac{GM}{r},\quad \mathbf{A}^\ast =\frac{GJ^\ast}{c}\frac{\hat{\mathbf{J}}\times \mathbf{r}}{r^3}\end{equation}
for $r>> GM/c^2$ and $r>>J^\ast /(Mc)$. Here $\hat{\mathbf{J}}$ is a fixed unit angular momentum vector of the source. With these potentials, Eq.~\eref{eq:7} is equivalent to the linear form of the Kerr metric in isotropic Cartesian coordinates.

It follows from the results of the previous section that within the GEM framework, metric \eref{eq:7} with
\begin{eqnarray}\label{eq:15} \Phi =\frac{GM}{r},\quad \mathbf{A}=\frac{GJ(t)}{c}\frac{\hat{\mathbf{J}} \times \mathbf{r}}{r^3},\\
\label{eq:16} J(t)=f(t)J^\ast \end{eqnarray}
represents the exterior field of a localized source of mass $M$ and angular momentem $J(t)$ that varies linearly with time along the fixed axis $\hat{\mathbf{J}}$. Substituting Eqs.~\eref{eq:15} and \eref{eq:16} in Eq.~\eref{eq:3}, one can find the {\em effective} source for these potentials; indeed,
\begin{equation}\label{eq:17} T_{00}=Mc^2\delta (\mathbf{r}),\quad T_{0i} =\frac{1}{2} c[\mathbf{J} (t)\times \boldsymbol{\nabla} ]_i\delta (\mathbf{r}).\end{equation}
Thus the mass-energy current is singular in this case and essentially confined to the origin of spatial coordinates. On the other hand, Eq.~\eref{eq:10} implies that
\begin{equation}\label{eq:18} T_{ij} =-\frac{1}{8\pi} \left[ (\dot{\mathbf{J}}\times \boldsymbol{\nabla} )_i \nabla_j \frac{1}{r} +(\dot{\mathbf{J}} \times \boldsymbol{\nabla} )_j \nabla _i \frac{1}{r}\right] ,\end{equation}
where $\dot{\mathbf{J}}=d\mathbf{J} /dt$ is a constant vector. The stresses \eref{eq:18} are time-independent and extend throughout space, but fall off to zero as $r^{-3}$ for $r\to \infty$. As emphasized before, the influence of the stresses on the spacetime metric is of $O(c^{-4})$ and is therefore ignored in the linear perturbation approach to GEM.

The localized mass-energy current that generates exterior GEM potentials \eref{eq:15} in an astrophysical situation could be considered as a variant of Newton's experiment with the rotating bucket of water but on a vastly different scale. Imagine an initially stationary configuration with angular momentum $J_I$. At some initial time $t_I$, the system undergoes a linear spin-up (or spin-down) that ends at $t_F$ and the system then returns to a stationary state for $t>t_F$ with angular momentum $J_F$. The total mass-energy of the configuration remains the same throughout. Thus Eq.~\eref{eq:15} holds for all time and $J$ is constant except for $t\in [t_I,t_F]$, where
\begin{equation}\label{eq:19} J=J_I +\frac{J_F-J_I}{t_F-t_I}(t-t_I).\end{equation}
Moreover, $\dot{J}(t)$ is piecewise constant and hence 
\begin{equation}\label{eq:20} \ddot{J}=\frac{J_F-J_I}{t_F-t_I} [\delta (t-t_I)-\delta (t-t_F)].\end{equation}
This implies, via Eq.~\eref{eq:3}, that there are additional spatially extended but instantaneous currents at $t_I$ and $t_F$ responsible for respectively turning on and off the temporal variation of the source. That is, Eq.~\eref{eq:17} holds except that $T_{0i}$ is modified at $t_I$ and $t_F$,
\begin{equation}\label{eq:21} T_{0i} =-\frac{1}{8\pi c}\left( \frac{\ddot{\mathbf{J}} \times \mathbf{r}}{r^3} \right) _i +\frac{1}{2} c (\mathbf{J}\times \boldsymbol{\nabla})_i\delta (\mathbf{r}).\end{equation}

It is interesting to explore the influence of time-varying gravitomagnetic fields on some standard consequences of general relativity. In the following subsections, we employ the nonstationary linearized Kerr spacetime to discuss some aspects of spin precession and the motion of null rays; the motion of test particles will be treated in \cite{5}. For the sake of simplicity, we limit our considerations to temporal intervals that are within $[t_I,t_F]$ and are such that $2 |\mathbf{A}| << c^2$.

\subsection{Gyroscope precession}\label{s:3.1}

The purpose of this subsection is to provide an estimate of the influence of intrinsic variability of the source on gyro precession. A detailed treatment of gyro motion in the special nonstationary gravitational field under consideration here on the basis of the Mathisson-Papapetrou-Dixon equations is clearly beyond the scope of this work~\cite{6}. Instead, we concentrate on a free test gyro held fixed outside the source with potentials given in Eq.~\eref{eq:15}. It follows from a simple extension of previous results~\cite{6} that to linear order in the spin, the gyro precesses with frequency $\mathbf{B}/c$, where the gravitomagnetic field is given by
\begin{equation}\label{eq:22} \mathbf{B}=\frac{GJ(t)}{cr^3} [3 (\hat{\mathbf{J}}\cdot \hat{\mathbf{r}}) \hat{\mathbf{r}}-\hat{\mathbf{J}} ].\end{equation}
The resulting motion can be simply described via introduction of a new temporal coordinate $\tau$ given by $d\tau =f(t)dt$.

For a test gyro in orbit about the source, the motion of the gyro spin is more complex; nevertheless, it is possible to estimate the influence of a variable mass-energy current on the gyro spin. For the Earth, the secular increase in the period of proper rotation $P_\oplus$ is approximately linear in time and can be characterized by $\dot{P}_\oplus /P_\oplus \approx 3\times 10^{-10}$ per year. This implies that the influence of the secular decrease of the Earth's angular momentum on the GP-B experiment~\cite{7} is entirely negligible. Indeed, the magnitude of the corresponding net decrease in the strength of the gravitomagnetic field of the Earth over the course of the GP-B experiment is about eight orders of magnitude smaller than the projected experimental sensitivity of the GP-B~\cite{7}.

The spin-curvature force on the test gyro is given by
\begin{equation}\label{eq:23} F^\alpha =-\frac{c}{2} R^\alpha _{\;\; \beta \mu\nu} u^\beta S^{\mu\nu},\end{equation}
where $u^\beta$ is the four-velocity of the gyro~\cite{6}. It follows that in the linear GEM framework and to first order in spin, $F^0=0$ and
\begin{equation}\label{eq:24} c\mathbf{F}=-\nabla (\mathbf{S}\cdot \mathbf{B}).\end{equation}
Thus the classical spin-gravity coupling is recovered, except that now $\mathbf{B}$ is linearly dependent upon time. We expect that the same holds for intrinsic spin with time-dependent Hamiltonian $H=\mathbf{S}\cdot \mathbf{B}/c$; time-dependent spin-rotation coupling has been considered in~\cite{8}. It is straightforward to verify this coupling for photon spin via the rotation of plane of polarization of electromagnetic radiation propagating in the exterior field of the source. The general scheme given in~\cite{1} implies that the result is
\begin{equation}\label{eq:25} \theta =\frac{1}{c^2}\int\mathbf{B}\cdot d\mathbf{r},\end{equation}
where the integration is carried out along the path of the null ray.

Imagine an electromagnetic ray that propagates along the axis of rotation of the source. During the time that the ray travels from $r_1$ to $r_2$, the angular momentum of the source changes linearly from  $J_1$ to $J_2$ and the plane of polarization of the electromagnetic wave rotates by an angle $\theta$, where
\begin{equation}\label{eq:26} \theta =\frac{G}{c^3}\frac{(r_2-r_1)}{r_1r_2} \left( \frac{J_1}{r_1}+\frac{J_2}{r_2}\right).\end{equation}
The standard Skrotskii result is recovered for $J_1=J_2$ \cite{1}.

\subsection{Time delay}\label{s:3.2}

We next examine the gravitomagnetic time delay of null geodesic rays in our model. The results may be of astrophysical interest in pulsar timing, for instance, as pulsars generally lose angular momentum due to external electromagnetic braking torques.

Within the linear GEM framework, the time delay $\Delta_G$ of a null ray propagating from $P_1:(ct_1, \mathbf{r}_1)$ to $P_2 :(ct_2, \mathbf{r}_2)$ is
\begin{equation}\label{eq:27} t_2-t_1 =\frac{1}{c} |\mathbf{r}_2-\mathbf{r}_1|+\Delta _G,\end{equation}
where \cite{9}
\begin{equation}\label{eq:28} \Delta_G=\frac{1}{2c}\int^{P_2}_{P_1} \bar{h}_{\alpha \beta} k^\alpha k^\beta d\zeta.\end{equation}
Here $k^\alpha=dx^\alpha /d \zeta =(1,\hat{\mathbf{k}})$ is the constant tangent vector of the unperturbed null ray and the integration is performed along the straight line from $P_1$ to $P_2$. Thus $\Delta_G=\Delta_{GE}+\Delta_{GM}$, where  $\Delta_{GE}$ is the familiar Shapiro time delay and
\begin{equation}\label{eq:29} \Delta_{GM}=-\frac{2}{c^3}\int^{P_2}_{P_1}\mathbf{A}\cdot d\mathbf{r}\end{equation}
is the gravitomagnetic time delay~\cite{9}.

It is interesting to work out $\Delta_{GM}$ for the exterior field given by Eq.~\eref{eq:15}. The result is
\begin{equation}\label{eq:30} \Delta_{GM}=-\frac{2G}{c^4} \frac{\hat{\mathbf{J}}\cdot (\hat{\mathbf{r}}_1 \times \hat{\mathbf{r}}_2)}{1+\hat{\mathbf{r}}_1 \cdot \hat{\mathbf{r}}_2} \left( \frac{J_1}{r_1}+\frac{J_2}{r_2}\right),\end{equation}
where Eqs.~\eref{A3} and \eref{A4} of the Appendix have been used. When the ray propagates from $P_1$ to $P_2$, the angular momentum of the source varies linearly from $J_1 $ to $J_2$; for $J_1 =J_2$ however, Eq.~\eref{eq:30} reduces to formula (16) of \cite{9}.

The time delay of null rays in the presence of a cosmological constant has been similarly calculated in \cite{10}. This treatment has been erroneously criticized in a recent paper~\cite{11}. Contrary to the claim in~\cite{11}, the argument in~\cite{10}, as is evident from the context, was simply based on the mathematical fact that for a null ray in Schwarzschild-de Sitter spacetime in standard coordinates, the geometric path of the ray in space does not suffer an additional deflection away from a straight line due to the presence of a cosmological constant---see, for instance, Eq.~(9) of \cite{11}.

\subsection{Deflection}\label{s:3.3}

Finally, for a null geodesic ray with tangent vector $K^\mu$ that travels from $P_1$ to $P_2$, $K^\mu (P_2)-K^\mu (P_1)=\Sigma^\mu$, where
\begin{equation}\label{eq:31} \Sigma ^\mu =-\int^{P_2}_{P_1} \Gamma^\mu_{\alpha\beta} k^\alpha k^\beta d\zeta \end{equation}
follows from the integration of the null geodesic equation in the linear GEM scheme. Using the connection coefficients given in~\cite{3} for metric \eref{eq:7} with potentials \eref{eq:15}, it is straightforward to compute the integral in Eq.~\eref{eq:31} along the unperturbed path of the ray. Thus $\Sigma^\mu =(\Sigma^0,\boldsymbol{\Sigma})$ is given by
\begin{eqnarray}\label{eq:32} \Sigma^0 =-\frac{2GM}{c^2} (Q\mathcal{I}_0+\mathcal{I}_1 ) \nonumber\\
\qquad +\frac{6G}{c^3}(\mathbf{r}_1 \times \hat{\mathbf{k}}) \cdot [\mathbf{J}_1 (Q\mathcal{L}_0+\mathcal{L}_1 )+\frac{1}{c} \dot{\mathbf{J}} (Q\mathcal{L}_1+\mathcal{L}_2)],\\
 \boldsymbol{\Sigma} =-\frac{2GM}{c^2} (\mathbf{r}_1 -Q\hat{\mathbf{k}}) \mathcal{I}_0-\frac{2G}{c^3L} (\mathbf{r}_1 \times \mathbf{J}_2-\mathbf{r}_2 \times \mathbf{J}_1)\mathcal{I}_0\nonumber\\
\qquad + \frac{6G}{c^3} \mathbf{r}_1 \times \hat{\mathbf{k}}\; [\mathbf{J}_1 \cdot (\mathbf{r}_1 \mathcal{L}_0 +\hat{\mathbf{k}}\mathcal{L}_1 )+\frac{1}{c}\dot{\mathbf{J}}\cdot (\mathbf{r}_1 \mathcal{L}_1 +\hat{\mathbf{k}}\mathcal{L}_2 )].\label{eq:33}\end{eqnarray}
Here $Q=\mathbf{r}_1\cdot \hat{\mathbf{k}}$, $L=|\mathbf{r}_2-\mathbf{r}_1|$ and
\begin{equation}\label{eq:34} \hat{\mathbf{k}}=\frac{1}{L} (\mathbf{r}_2 -\mathbf{r}_1),\quad \dot{\mathbf{J}}=\frac{c}{L} (J_2-J_1)\hat{\mathbf{J}} .\end{equation}
The integrals $\mathcal{I}_0$, $\mathcal{I}_1$, $\mathcal{L}_0$, $\mathcal{L}_1$ and $\mathcal{L}_2$ are given in the Appendix. It is interesting to note that
\begin{equation}\label{eq:35} \boldsymbol{\Sigma} \cdot \hat{\mathbf{k}}=\frac{2G\dot{J}}{c^4} \mathbf{r}_1 \cdot (\hat{\mathbf{k}}\times \hat{\mathbf{J}})\mathcal{I}_0.\end{equation}

The result of this calculation may be of interest in connection with the significant advances that are expected to occur in microarcsecond astrometry in the near future~\cite{12}.

\section{Discussion}\label{s:4}

The gravitational field equations have solutions in which the mass-energy and angular momentum of the source vary with time as a consequence of emission or absorption of radiation. However, the present paper is devoted to nonradiative situations involving the slow temporal variation of the angular momentum of the source. A method is presented to estimate the influence of such variability on gravitomagnetic phenomena. 

The main result of this work is based on the observation that within the linear GEM framework it is possible to transform a localized stationary configuration to one with a gravitomagnetic field that varies linearly with time. Starting with the special nonstationary spacetime corresponding to the linearized Kerr metric in isotropic Cartesian coordinates, the consequences of the linear temporal variation of the gravitomagnetic vector potential are explored in some situations of physical interest such as the GP-B experiment, the rotation of plane of polarization of electromagnetic waves and the time delay as well as deflection of electromagnetic signals. However, a proper discussion of
the methods needed to deduce observable effects from the results presented
here is beyond the scope of this work.

\appendix \setcounter{section}{1}
\section*{Appendix}

The purpose of this appendix is to evaluate the integrals
\begin{equation}\label{A1} \mathcal{I}_n=\int^L_0\frac{\zeta^nd\zeta}{r^3},\quad \mathcal{L}_m=\int^L_0 \frac{\zeta ^md\zeta}{r^5}\end{equation}
for $n=0,1$ and $m=0,1,2$. Here
\begin{equation}\label{A2} \mathbf{r}(\zeta )=\mathbf{r}_1 +\hat{\mathbf{k}}\zeta ,\end{equation}
while $\hat{\mathbf{k}}$ and $L$ have been defined in section~\ref{s:3}. These integrals can be computed using formulas 2.263 and 2.264 given respectively on pages 82 and 83 of~\cite{13}. We find that
\begin{equation}\label{A3} \mathcal{I}_0 =\frac{1}{\delta} \hat{\mathbf{k}} \cdot (\hat{\mathbf{r}}_2-\hat{\mathbf{r}}_1 ),\quad \mathcal{I}_1 =\frac{r_1}{\delta} (1-\hat{\mathbf{r}}_1 \cdot \hat{\mathbf{r}}_2),\end{equation}
where $\delta =r^2_1 -Q^2$, $Q=\hat{\mathbf{k}}\cdot \mathbf{r}_1$ as before, and hence
\begin{equation}\label{A4} \delta L^2=r^2_1 r^2_2 [1-(\hat{\mathbf{r}}_1 \cdot \hat{\mathbf{r}}_2)^2].\end{equation}
Furthermore,
\begin{eqnarray}\label{A5} \mathcal{L}_0 &=&\frac{1}{3\delta } \left(2\mathcal{I}_0+\frac{L+Q}{r^3_2} -\frac{Q}{r^3_1}\right),\\
\label{A6}\mathcal{L}_1 &=& -Q\mathcal{L}_0 -\frac{1}{3} \left(\frac{1}{r^3_2}-\frac{1}{r^3_1}\right),\\
\label{A7} \mathcal{L}_2 &=& \frac{1}{2} r^2_1 \mathcal{L}_0 -\frac{1}{2} Q\mathcal{L}_1 -\frac{1}{2} \frac{L}{r^3_2}.\end{eqnarray}

\end{document}